\documentclass[twocolumn,pre,twoside,superscriptaddress,floatfix,showkeys,showpacs]{revtex4}

\usepackage{amsmath}
\usepackage{amssymb}
\usepackage[usenames]{color}
\usepackage{graphicx}
\usepackage{dcolumn}
\usepackage{bm}

\def\dps{\displaystyle}
\def\m#1{\mathrm{#1}}

\def\d{\mathrm{d}}

\def\epsilon{\varepsilon}
\def\theta{\vartheta}
\def\rho{\varrho}

\def\Re{\mathop\mathrm{Re}}

\begin{document}


\title{Salt-induced microheterogeneities in binary liquid mixtures}

\author{Markus Bier}
\email{bier@is.mpg.de}
\affiliation
{
   Max Planck Institute for Intelligent Systems,
   Heisenbergstr.\ 3,
   70569 Stuttgart,
   Germany
}
\affiliation
{
   Institute for Theoretical Physics IV,
   University of Stuttgart,
   Pfaffenwaldring 57,
   70569 Stuttgart,
   Germany
}

\author{Julian Mars}
\affiliation
{
   Max Planck Institute for Polymer Research, 
   Ackermannweg 10, 
   55128 Mainz, 
   Germany
}
\affiliation
{
   Institute of Physics and
   MAINZ Graduate School, 
   Johannes Gutenberg University Mainz, 
   Staudingerweg 7,
   55128 Mainz, 
   Germany
}

\author{Hailong Li}
\affiliation
{
   Max Planck Institute for Polymer Research, 
   Ackermannweg 10, 
   55128 Mainz, 
   Germany
}

\author{Markus Mezger}
\affiliation
{
   Institute of Physics and
   MAINZ Graduate School, 
   Johannes Gutenberg University Mainz, 
   Staudingerweg 7,
   55128 Mainz, 
   Germany
}
\affiliation
{
   Max Planck Institute for Polymer Research, 
   Ackermannweg 10, 
   55128 Mainz, 
   Germany
}

\date{3 July 2017}

\begin{abstract}
The salt-induced microheterogeneity (MH) formation in binary liquid mixtures is studied by 
small-angle X-ray scattering (SAXS) and liquid state theory. 
Previous experiments have shown that this phenomenon occurs for antagonistic salts, whose cations
and anions prefer different components of the solvent mixture.
However, so far the precise mechanism leading to the characteristic length scale of MHs remained
unclear.
Here, it is shown that MHs can be generated by the competition of short-ranged interactions
and long-ranged monopole-dipole interactions.
The experimental SAXS patterns can be quantitatively reproduced by fitting to the derived 
correlation functions without assuming any specific model.
The dependency of the MH structure with respect to ionic strength and temperature is analyzed.
Close to the demixing phase transition, critical-like behavior occurs with respect to the spinodal
line in the phase diagram.
\end{abstract}


\maketitle


\section{Introduction}

Structure formation in the bulk of some complex fluids is a well known phenomenon.
Examples include the self-assembly of 
amphiphiles, block copolymers, room-temperature ionic liquids, or ionic surfactants into
micelles, microemulsions, lyotropic phases, or other microscopic heterogeneities 
\cite{Moroi1992, Zhang2003, Neto2005, Lazzari2006, Mason2006, AlShamery2008, Stubenrauch2009,
Mueller2011, Bier2017}.
There, the structure formation can be easily understood in terms of head-tail asymmetries of the 
composing molecules or in terms of an asymmetry generated by external fields 
\cite{Tsori2007}.
Commonly, the different components of the system or molecular groups can be classified in terms of
their hydrophilic/hydrophobic or polar/apolar character.
In most cases, the structural length scales of the systems are then governed by the specific
molecular dimensions of these molecular moieties.

However, there are complex fluids that become heterogeneous on length scales well above the 
molecular dimensions.
These fluids comprise binary liquid mixtures in the presence of antagonistic salts, 
i.e., systems where cations and anions are preferentially dissolved in different
components.
Experimentally, indications for microheterogeneity (MH) formation occurred by means of light
and small-angle X-ray scattering (SAXS).
An additional length scale was first observed in water, 3-methylpyridine, and sodium bromide
($\m{NaBr}$) mixtures \cite{Jacob1998,Jacob1999,Anisimov2000}.
A peculiarity of this system is the possible existence of a tricritical point. 
These studies led to some controversies, which have been resolved by realizing that the postulated
MHs were non-equilibrium structures with a long relaxation time \cite{Kostko2004,Wagner2004,
Leys2013}. 
Later, small-angle neutron scattering (SANS) provided clear evidence for equilibrium MHs in
mixtures of water, 3-methylpyridine, and sodium tetraphenylborate ($\m{NaBPh_4}$) 
\cite{Sadakane2007b,Sadakane2009,Sadakane2011,Sadakane2013,Sadakane2014}.
In contrast, no pronounced MH could be found for mixtures with inorganic salts 
\cite{Takamuku2001a,Takamuku2001b,Sadakane2006,Sadakane2007a,Takamuku2007a,Takamuku2007b,
Takamuku2007c,Haramaki2013}.

So far, theoretical studies addressing the MH formation have been concentrated on models
comprised of a single solvent component \cite{Nabutovskii1980,Nabutovskii1985,Bier2012b}
or a binary mixture solvent in the incompressibility limit \cite{Onuki2004,Onuki2011a,Onuki2011b,
Bier2012a,Onuki2016} with dissolved anions and cations.
This approach neglects the binary character of the solvent mixture.
Thus, MH formation is governed solely by the solvation contrast of the ion species in the solvent.
This led to an interpretation where the antagonistic salt ions behave similar to ionic surfactants
\cite{Sadakane2011,Sadakane2013,Sadakane2014}.
Recently, this generic approach found some support by SAXS measurements on water, 
2,6-dimethylpyridine, and quaternary ammonium bromide salt mixtures \cite{Witala2016}.
However, there are other systems where attributing MH formation to solvation contrasts
alone \cite{Bier2012b} does not apply \cite{Sadakane2011,Sadakane2013,Sadakane2014}.
Hence, full understanding of the underlying MH formation mechanisms is still to be achieved.

\begin{figure}[!t]
   \includegraphics{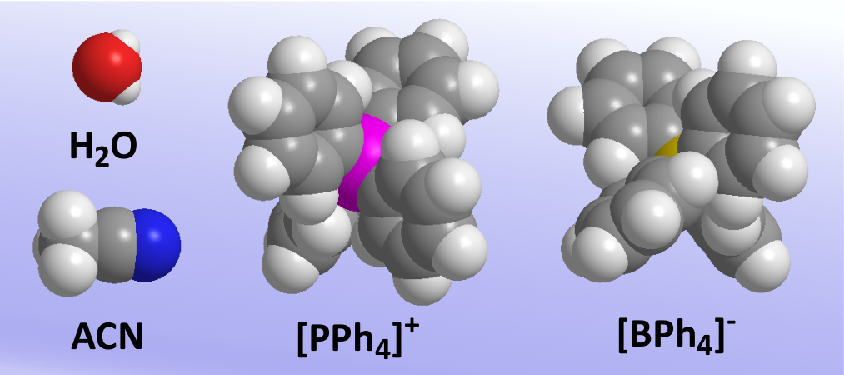}
   \caption{Molecular structure of the solvent and antagonistic salt components: water 
            ($\m{H_2O}$), acetonitrile (ACN), $\m{[PPh_4]^+}$ cations, and $\m{[BPh_4]^-}$ anions.
            Color code: oxygen (red), nitrogen (blue), phosphorus (pink), boron (yellow), 
            carbon (grey), and hydrogen (light grey).}
   \label{fig:molecules}
\end{figure}

The present work concentrates on an important aspect of this general problem, namely the question
about the origin of a characteristic length scale of the MH.
Experience tells that characteristic length scales are typically the result of competing
mechanisms, and it is the present goal to identify these for a particular type of systems.
Here, mixtures of water ($\m{H_2O}$), acetonitrile ($\m{ACN}$, $\m{CH_3CN}$), and the
antagonistic salts $\m{NaBPh_4}$ or tetraphenylphosphonium chloride ($\m{PPh_4Cl}$) 
(Sec.~\ref{Subsec:Setting}, Fig.~\ref{fig:molecules}) are studied by means of SAXS 
(Sec.~\ref{Subsec:SAXS}).
The resulting scattering intensity is analyzed in terms of a novel generic form derived in 
Sec.~\ref{Subsec:GenFormScatIn}.
In contrast to previous treatments \cite{Nabutovskii1980,Nabutovskii1985,Onuki2004,Onuki2011a,
Onuki2011b,Bier2012a,Bier2012b}, which are based on the intuitive picture of MH
formation being generated by long-ranged monopole-monopole interactions between the ions, here
long-ranged monopole-dipole interactions are taken into account.
Indeed, it is found for the present system that Coulomb interactions between the ions alone
cannot account for MH formation, but that monopole-dipole interactions between ions and solvent
molecules are decisive for the MH formation.
In Sec.~\ref{Sec:ResultsDiscussion} first the SAXS data are discussed (Sec.~\ref{Subsec:Fits}).
Next, the concept of a spinodal line is introduced (Sec.~\ref{Subsec:Spinodal}) and the
critical-like behavior of the system with respect to this spinodal line is verified 
(Sec.~\ref{Subsec:Critical}).
Moreover, the dependence of the characteristic length scale of MHs on the temperature and
the ionic strength are determined (Sec.~\ref{Subsec:StrucMH}).
Finally, based on the proposed approach accounting for long-ranged monopole-dipole interactions,
competing mechanisms, which can give rise to the characteristic length scale of MHs, are
propounded in Sec.~\ref{Sec:conclusion_summary}.

\section{\label{Sec:methods}Method and Experiments}


\subsection{\label{Subsec:Setting}Setting}

The systems under consideration are binary mixtures of polar solvents, denoted as components 
``$A$'' and ``$B$''.
The mixture exhibits a miscibility gap with an upper or lower critical demixing point at mole 
fraction $x_\m{A,c}$ of component $A$ and temperature $T_\m{c}$.
An extensive summary of binary solvent mixtures with lower (LCST) and upper (UCST) critical
points were compiled by Francis \cite{Francis1961}.
In this liquid, univalent ions of an antagonistic salt composed of cations ``$\oplus$'' and anions
``$\ominus$'' are dissolved.
To experimentally study the MH close to the critical demixing point $\left( T_\m{c}, x_\m{A,c} 
\right)$ by SAXS, the system ideally fulfills a series of requirements: 

(1) The critical temperature $T_c$ should be in the experimentally accessible
temperature range and $x_\m{A,c} \approx 1/2$.

(2) Around room temperature, both components should be miscible in all proportions. 

(3) The two components $A$ and $B$ of the mixture should exhibit a good X-ray
scattering contrast.
In forward direction, the scattering contrast is quantified by the difference of the real 
parts $\Re\left(n\right) = 1 - \delta$ of the refractive indices of the two solvent components.
For hard X-rays of energy $E$ and soft matter composed of elements from the first or
second period, the refractive index decrement $\dps\delta \approx 0.23\cdot10^{-3}
\frac{\m{cm^3keV^2}}{\m{g}}\,\frac{\varrho_\m{m}}{E^2}$ can be estimated from the mass density 
$\varrho_\m{m}$ \cite{Henke1993}. 

\begin{table}[!t]
   \begin{tabular}{|l||c|c|c|c|c|}
      \hline
                 & $T_\m{m}$      & $\rho_\m{m}$     & $\delta$  & $p$     & $\varepsilon_\text{r}$ \\
                 & $\m{^\circ C}$ & $\m{g\,cm^{-3}}$ & $10^{-6}$ & $\m{D}$ &                        \\
      \hline
      $\m{H_2O}$ & $0.0026$       & $0.9982$         & $3.58$    & $1.855$ & 80                     \\
      $\m{ACN}$  & $-44$          & $0.7857$         & $2.72$    & $3.925$ & 36                     \\
      $\m{EDO}$  & $11.75$        & $1.0337$         & $3.63$    & $<0.4$  & 2.2                    \\
      $\m{3MP}$  & $-18.1$        & $0.9566$         & $3.31$    & $2.4$   & 10                     \\
      $\m{DMP}$  & $-6.12$        & $0.9226$         & $3.22$    & $1.66$  & 6.9                    \\
      \hline
   \end{tabular}
   \caption{Solvent properties of water ($\m{H_2O}$), acetonitrile (ACN), 1,4-dioxane (EDO), 
            3-methylpyridine (3MP), and 2,6-dimethylpyridine (DMP): melting point $T_\m{m}$
            \cite{Haynes2017}, mass density $\rho_\m{m}$ \cite{Haynes2017}, refractive index
            decrement $\delta$ for $8\,\m{keV}$ X-rays \cite{Henke1993}, electric dipole
            moment $p$ \cite{Williams1930, Haynes2017}, and static dielectric constant 
            $\varepsilon_\text{r}$ \cite{Wohlfarth2008}.}
   \label{Tab:solvents}
\end{table}

(4) The solubility of the antagonistic salt in the solvent mixture should be
$\gtrsim 100\,\m{mM}$.
Table~\ref{Tab:solvents} summarizes the relevant parameters of solvents in which MHs have
been previously studied experimentally.


\subsection{\label{Subsec:System}Material system}

For the  experiments presented in this work mixtures of water ($\m{H_2O}$) and acetonitrile 
($\m{ACN}$) were studied by SAXS (Sec.~\ref{Subsec:SAXS}).
The system exhibits a miscibility gap with an upper critical demixing point at $x_\m{H_2O,c}=0.638$
and $T_\m{c} = -1.34\,\m{^\circ C}$ \cite{Szydlowski1999}.
Comparison of the $\m{H_2O}$ dipole moment ($1.855\,\m{D}$) with $\m{ACN}$ ($3.925\,\m{D}$), 
renders $\m{ACN}$ the more polar component.
At $8\,\m{keV}$, the refractive index decrement $\delta$ for $\m{H_2O}$ 
($\delta = 3.58 \cdot 10^{-6}$) is 32\% larger than for $\m{ACN}$ ($\delta = 2.72 \cdot 10^{-6}$).
Therefore, compared to water/3MP mixtures used in previous studies \cite{Sadakane2006,Sadakane2007a,
Sadakane2007b,Sadakane2011,Sadakane2014}, the water/ACN system provides a much larger scattering contrast
in SAXS experiments (Tab.~\ref{Tab:solvents}).
Here, salts with the cations $\m{Na^+}$ or $\m{[PPh_4]^+}$ and with the anions $\m{Cl^-}$ or
$\m{[BPh_4]^-}$ were studied.
From the Gibbs free energies of transfer $\Delta_\m{t}G^\circ(\m{H_2O}\to\m{ACN})$ it is inferred
that $\m{Na^+}$ and $\m{Cl^-}$ ions prefer $\m{H_2O}$ over $\m{ACN}$, whereas $\m{[BPh_4]^-}$ and 
$\m{[PPh_4]^+}$ ions prefer $\m{ACN}$ over $\m{H_2O}$ \cite{Marcus1983,Inerowicz1994,Kalidas2000}.
Thus, $\m{NaBPh_4}$ and $\m{PPh_4Cl}$ can be considered as antagonistic salts.
To verify the importance of the antagonistic character of the salt for MH formation, 
$\m{NaCl}$ with a solubility of $6.1 \, \mathrm{M}$ in $\m{H_2O}$ and $40 \, \mu\mathrm{M}$ in ACN
served as an example for a hydrophilic salt \cite{Burgess1978}.
In contrast, $\m{[PPh_4][BPh_4]}$ has a solubility of $2.72 \, \m{n}\mathrm{M}$ in $\m{H_2O}$ and
$1.14\, \mathrm{mM}$ in ACN \cite{Popovych1981}.
However, its solubility in the mixtures was too low to experimentally study the presence of MH.
Measurements were performed for $\m{H_2O}$ mole fractions $x_\m{H_2O}\in\{0.635,0.7,0.8\}$ and
ionic strengths $I\in\{10,50,200\}\,\m{mM}$.
Purified water was prepared by ultrafiltration and deionization (Sartorius Arium 611 VF, 
$18.2\,\m{M\Omega}$).
Other chemicals, ACN (Fisher Chemicals, HPLC grade), $\m{NaBPh_4}$ (Sigma Aldrich, $\geq 99.5 \%$),
$\m{PPh_4Cl}$ (Sigma Aldrich, $\geq 98.0 \%$), and $\m{NaCl}$ (Sigma Aldrich, $\geq 99.8 \%$) were
used as received.
Robustness of the results has been verified by repeated preparation and SAXS measurements of
some compositions.


\subsection{\label{Subsec:SAXS}Small-angle X-ray scattering}

SAXS measurements were performed at a self-constructed instrument \cite{Weiss2017} using a rotating
$\m{Cu}$ anode X-ray generator (Rigaku MicroMax 007). 
The beam was monochromatized (wavelength $\lambda=1.54\,\m{\AA}$) and collimated by a 
multilayer optic (Osmic Confocal Max-Flux, $\m{Cu}\ K_\alpha$) and three 4-jaw slit sets 
($500\times500\,\m{\mu m^2}$ slit gap) with $150\,\m{cm}$ collimation length. 
An incident X-ray flux of approx.\ $10^7\,\m{photons/s}$ at the sample position was measured by an
inversion layer silicon photodiode (XUV-100, OSI Optoelectronics). 
Samples were contained in $1\,\m{mm}$ thick sealed glass capillaries, placed in a temperature
controlled holder (stability better $\pm0.05\,\m{K}$), and mounted inside the vacuum chamber. 
2D diffraction patterns were recorded on an online image plate detector (Mar345). 
The sample-detector distance of $210\,\m{cm}$ was calibrated with silver behenate \cite{Huang1993,
Gilles1998}.
SAXS data, collected during three or more independent measurements with $1200\,\m{s}$ exposure time
each, were averaged and corrected by dark images. 
Artifacts, originating from high energy radiation, were removed by differential Laplace filtering. 
By azimuthal integration, the 2D data sets were converted to scattering intensities 
$\mathcal{I}\left(q\right)$ vs.\ momentum transfer $q = 4\pi/\lambda\ \sin(\vartheta)$ with
total scattering angle $2\vartheta$.
To focus on the scattering from MHs, for all data sets the corresponding scattering 
patterns recorded at $25\,\m{^\circ C}$ and constant, $q$-independent 
offset values were subtracted from the raw data.
This ensures that for sufficiently high $q$-values the average intensity 
$\langle\mathcal{I}\left(q\right) \rangle$ vanishes.


\subsection{\label{Subsec:GenFormScatIn}Generic form of the scattering intensity}

In order to gain physical insight from the measured scattering intensities 
$\mathcal{I}\left(q\right)$, a fitting function is required which allows for an interpretation of
the underlying model parameters.
The derivation of the fitting function used in the present work is based on a model-free reasoning
in terms of the direct correlation functions $c_{ij}(r), i,j\in\{A,B,\oplus,\ominus\}$.
This approach is similar to the one employed in Ref.~\cite{Bier2012b}.

In a first step one splits the 3D-Fourier integrals
\begin{align}
   \widehat{c}_{ij}(q) = \frac{4\pi}{q}\int\limits_0^\infty{\d} r\;rc_{ij}(r)\sin(qr) 
                       = \widehat{c}^<_{ij}(q) + \widehat{c}^>_{ij}(q)
   \label{eq:chat}
\end{align}
of the direct correlation functions $c_{ij}(r)$ \cite{Hansen1986} with 
\begin{align}
   \widehat{c}^<_{ij}(q) &= \frac{4\pi}{q}\int\limits_0^R{\d} r\;rc_{ij}(r)\sin(qr), 
   \label{eq:chatlow}\\
   \widehat{c}^>_{ij}(q) &= \frac{4\pi}{q}\int\limits_R^\infty{\d} r\;rc_{ij}(r)\sin(qr). 
   \label{eq:chatup}
\end{align}

As the integration range in Eq.~(\ref{eq:chatlow}) is a compact interval for any finite value $R$,
$\widehat{c}^<_{ij}(q)$ is an even and entire function, i.e., it possesses an expansion of the form
\begin{align}
   \widehat{c}^<_{ij}(q) = c^{<(0)}_{ij} + c^{<(2)}_{ij}q^2 + \mathcal{O}(q^4).
   \label{eq:clowexp} 
\end{align}
Short-ranged interactions, e.g.,  due to solvation, formation of coordination complexes, 
or hydrogen bonding, contribute only to this part of the direct correlation function,
provided the range $R$ is larger than the interaction range.

For sufficiently large $R$, the direct correlation functions are given by $c_{ij}(r)\simeq
-\beta U_{ij}(r)$ at distances $r>R$ with the pair interaction potential $U_{ij}(r)$ of species
$i$ and $j$ \cite{Hansen1986}.
As solvent molecules are electrically neutral, i.e., they do not carry an electric 
monopole, only dipole-dipole interactions are present asymptotically, i.e., 
$U_{ij}(r>R) \simeq A^{(6)}_{ij}/r^6$ for $i,j\in\{A,B\}$.
Note that here ``dipole'' refers to permanent, induced, or spontaneous dipoles and that 
permanent dipoles are orientationally disordered.
In contrast, the asymptotic interactions at long distances $r>R$ between a solvent molecule
and an ion, which by definition carries an electric monopole, are not only of the type 
monopole-dipole, but additional dipole-dipole contributions (Van der Waals forces) occur, i.e., 
$U_{ij}(r>R) \simeq A^{(4)}_{ij}/r^4 + A^{(6)}_{ij}/r^6$ for $i\in\{A,B\},j\in \{\oplus,\ominus\}$.
Similarly, two ions, both of which carry electric monopoles, interact asymptotically at
long distances $r>R$ with monopole-monopole, monopole-dipole, and
dipole-dipole contributions, i.e., $U_{ij}(r>R) \simeq A^{(1)}_{ij}/r + A^{(4)}_{ij}/r^4 + 
A^{(6)}_{ij}/r^6$ for $i,j\in\{\oplus,\ominus\}$ with $A^{(1)}_{ij}=z_iz_j\ell_B/\beta$, 
$z_\oplus=1$, $z_\ominus=-1$ and the Bjerrum length $\ell_B=\beta e^2/(4\pi\varepsilon_0
\varepsilon)$.
 
A straightforward expansion of
\begin{align}
   W(\alpha) := \frac{4\pi}{q}\int\limits_R^\infty{\d} r\;r^{1-\alpha}\sin(qr)
   \label{eq:Wdef}
\end{align}
in powers of $q$ leads to \cite{Gradshteyn}
\begin{align}
   W(1) &= 4\pi\left(\frac{1}{q^2} -\frac{R^2}{2} + \frac{R^4}{24}q^2 + \mathcal{O}(q^4)\right) 
   \label{eq:W1}\\
   W(4) &= 4\pi\left(\frac{1}{R} - \frac{\pi}{2}q + \frac{R}{6}q^2 + \mathcal{O}(q^4)\right) 
   \label{eq:W4}\\
   W(6) &= 4\pi\left(\frac{1}{3R^3} - \frac{1}{6R}q^2 + \mathcal{O}(q^3)\right).
   \label{eq:W6}
\end{align}
From Eq.~(\ref{eq:chatup}) one infers
\begin{align}
   c^>_{ij}(q) = -\beta A^{(6)}_{ij}W(6)
   \label{eq:cupsolsol}
\end{align}
for $i,j\in\{A,B\}$,
\begin{align}
   c^>_{ij}(q) = -\beta\left(A^{(4)}_{ij}W(4) + A^{(6)}_{ij}W(6)\right)
   \label{eq:cupsolion}
\end{align}
for $i\in\{A,B\},j\in\{\oplus,\ominus\}$, and
\begin{align}
   c^>_{ij}(q) = -\beta\left(A^{(1)}_{ij}W(1) + A^{(4)}_{ij}W(4) + A^{(6)}_{ij}W(6)\right)
   \label{eq:cupionion}
\end{align}
for $i,j\in\{\oplus,\ominus\}$.

Combining the expansions in Eqs.~(\ref{eq:clowexp}) and 
(\ref{eq:cupsolsol})--(\ref{eq:cupionion}) one obtains
from Eq.~(\ref{eq:chat}) the expansions
\begin{align}
   \widehat{c}_{ij}(q) = 
   c^{(0)}_{ij} + c^{(2)}_{ij}q^2 + \mathcal{O}(q^3)
   \label{eq:chatsolsol}
\end{align}
for $i,j,\in\{A,B\}$,
\begin{align}
   \widehat{c}_{i,j}(q) = 
   c^{(0)}_{ij} + c^{(1)}_{ij}q + c^{(2)}_{ij}q^2 + \mathcal{O}(q^3)
   \label{eq:chatsolion}
\end{align}
for $i,\in\{A,B\},j\in\{\oplus,\ominus\}$, and
\begin{align}
   \widehat{c}_{ij}(q) = 
   -\frac{z_iz_j\ell_B}{q^2} + c^{(0)}_{ij} + c^{(1)}_{ij}q + c^{(2)}_{ij}q^2 + \mathcal{O}(q^3)
   \label{eq:chationion}
\end{align}
for $i,j\in\{\oplus,\ominus\}$.

The coefficients $c^{(k)}_{ij}$ depend on the system as well as
on the thermodynamic state.
Note that, due to Eq.~(\ref{eq:W4}), non-vanishing coefficients $c^{(1)}_{ij}\not=0$ can occur
only in the presence of long-ranged monopole-dipole interactions.

In order to calculate the partial structure factors, the $4\times4$-matrix 
$\underline{\underline{\mathcal{C}}}$ with components $\mathcal{C}_{ij}:=
\sqrt{\varrho_i\varrho_j}\,\widehat{c}_{ij}(q)$ is introduced.
Here, $\varrho_i$ is the bulk number density of species $i$.
Then, one obtains the matrix $\underline{\underline{\mathcal{S}}}=
(\underline{\underline{1}}-\underline{\underline{\mathcal{C}}})^{-1}$, whose
components  $\mathcal{S}_{ij}$ are related to the partial structure factors $S_{ij}(q) = 
\sqrt{\varrho_i\varrho_j}\mathcal{S}_{ij}/\varrho$, where $\varrho=\sum_i\varrho_i$ denotes the
total number density \cite{Hansen1986}.
Here, only wave numbers $q$ corresponding to length scales larger than the molecular sizes are 
considered, where the form factors of the solvent species $i\in\{A,B\}$ are essentially given by
the numbers $Z_i$ of electrons per molecule: $\dps\mathcal{I}\left(q\right)\sim 
\sum_{i,j\in\{A,B\}} Z_iZ_jS_{ij}(q)$.
Performing the matrix inversion in $\underline{\underline{\mathcal{S}}}=
(\underline{\underline{1}}-\underline{\underline{\mathcal{C}}})^{-1}$ by means of Cramer's rule 
one obtains the following Pad\'{e} approximation of the scattering intensity in the range of large
length scales ($q\to0$):
\begin{align}
   \mathcal{I}\left(q\right)\simeq\frac{aq^2 + bq + c}{q^2 + mq + n}.
   \label{eq:Ipade}
\end{align}
The coefficients $a$, $b$, $c$, $m$, and $n$ in Eq.~(\ref{eq:Ipade}) depend on the 
system and on the thermodynamic state.
Given any specific model for the system under consideration, one would obtain explicit
expressions of these coefficients.
However, within the general approach of the present work, one can merely expect the
coefficients $a$, $c$, and $n$ in Eq.~(\ref{eq:Ipade}) to be positive.
For later reference the expressions of coefficients $m$ and $n$ are given in the form
\begin{widetext}
\begin{align}
   m = -M\left(
   c^{(1)}_{\oplus\oplus}+2c^{(1)}_{\oplus\ominus}+c^{(1)}_{\ominus\ominus} 
   + 2\frac{\big(1-c^{(0)}_{AA}\big)T^{(0)}_BT^{(1)}_B + c^{(0)}_{AB}\big(T^{(0)}_AT^{(1)}_B+
   T^{(0)}_BT^{(1)}_A\big) + \big(1-c^{(0)}_{BB}\big)T^{(0)}_AT^{(1)}_A}{\big(1-c^{(0)}_{AA}\big)
   \big(1-c^{(0)}_{BB}\big) - \big(c^{(0)}_{AB}\big)^2}\right)
   \label{eq:m}
\end{align}
\end{widetext}
with $T^{(k)}_i:=c^{(k)}_{i\oplus}+c^{(k)}_{i\ominus}$ and
\begin{align}
   n = N\left(1+\frac{I\bar{n}}{\big(1-c^{(0)}_{AA}\big)
   \big(1-c^{(0)}_{BB}\big) - \big(c^{(0)}_{AB}\big)^2}\right).
   \label{eq:n}
\end{align}
The positive coefficientes $M$, $N$, and $\bar{n}$ in Eqs.~(\ref{eq:m}) and (\ref{eq:n}) are 
system- and state-dependent.
Moreover, $M$ and $N$ vanish in the salt-free case ($I=0$). 
Writing the denominator in Eq.~(\ref{eq:Ipade}) in the form $(q+m/2)^2+n-m^2/4$, one recognizes
for $m<0$ the occurrence of a maximum of $\mathcal{I}\left(q\right)$ at $q=q_\m{max}:=-m/2$ with
a peak width of half height $2/\xi$, where $\xi:=1/\sqrt{n-m^2/4}$.


\section{\label{Sec:ResultsDiscussion}Results and Discussion}


\subsection{\label{Subsec:Fits}Fits of the scattering intensity} 

\begin{figure}[!t]
   \includegraphics{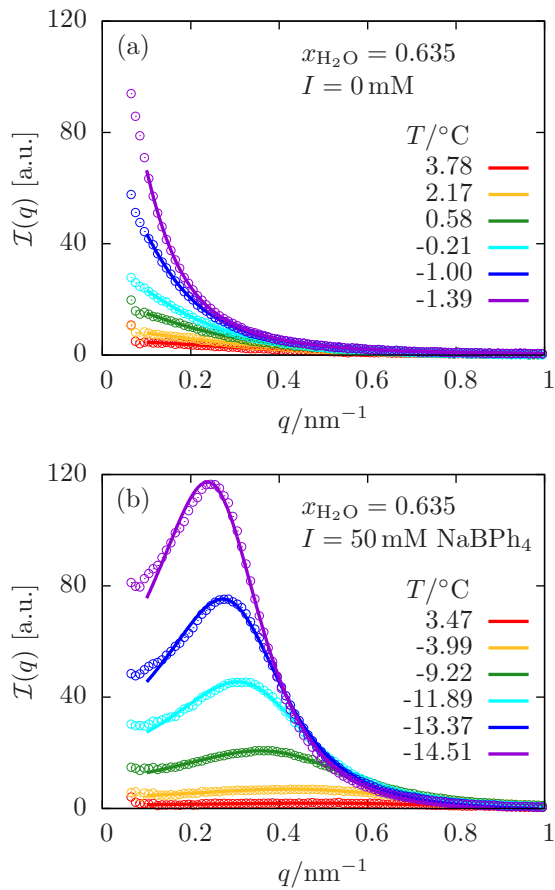}
   \caption{Scattering intensities $\mathcal{I}\left(q\right)$ of binary mixtures of $\m{H_2O}$
            and $\m{ACN}$ for water mole fraction $x_\m{H_2O}=0.635$ at various temperatures $T$.
            Circles represent the measured SAXS data, whereas lines correspond to fits of 
            Eq.~(\ref{eq:Ipade}).
            Panel (a) displays the case of the pure, salt-free mixture, which leads to monotonically
            decaying $\mathcal{I}\left(q\right)$ with a maximum at $q=0$.
            Panel (b) shows the case of $I=50\,\m{mM}$ $\m{NaBPh_4}$ added to the mixture in panel
            (a), which exhibits a maximum of $\mathcal{I}\left(q\right)$ at wave number 
            $q=q_\m{max}>0$.
            In both panels (a) and (b) the height of the maxima increases upon decreasing the 
            temperature, i.e., upon approaching the two-phase coexistence region in the phase
            diagram.
            Moreover, panel (b) shows a decrease of $q_\m{max}$, i.e., an increase of the length
            scale $2\pi/q_\m{max}$ of the MH, upon decreaseing the temperature.}
   \label{Fig:1}
\end{figure}

Figure~\ref{Fig:1} displays examples of measured scattering intensities $\mathcal{I}\left(q\right)$
(circles) and the corresponding fits according to Eq.~(\ref{eq:Ipade}) (lines) for water mole
fraction $x_\m{H_2O}=0.635$ at various temperatures $T$. 
In Fig.~\ref{Fig:1}(a) the case of a pure, salt-free ($I=0$) mixture is shown, where 
$\mathcal{I}\left(q\right)$ is monotonically decreasing with a maximum at wave number $q=0$.
The increase of the maximum $\mathcal{I}(0)$ upon decreasing the temperature $T$ is related to the
approach of the critical point at $x_{\m{H_2O},c}=0.638,T_c=-1.34\,\m{^\circ C}$ 
(Sec.~\ref{Subsec:System}).
Qualitatively the same monotonically decreasing scattering intensities $\mathcal{I}\left(q\right)$
have been observed for all pure, salt-free mixtures as well as for the mixtures with added 
$\m{NaCl}$.

In contrast, adding one of the antagonistic salts $\m{NaBPh_4}$ or $\m{PPh_4Cl}$ to an
$\m{H_2O}$/$\m{ACN}$ mixture results in non-monotonic scattering intensities 
$\mathcal{I}\left(q\right)$, as is displayed in Fig.~\ref{Fig:1}(b) for $x_\m{H_2O}=0.635$ with 
$I=50\,\m{mM}$ $\m{NaBPh_4}$.
Upon decreasing the temperature $T$, the height of the maxima $\mathcal{I}(q_\m{max})$ increases
and the wave numbers $q_\m{max}$ of the maximum shift towards smaller values.
These properties are discussed more systematically in the following sections.
The conclusion here is that the occurrence of a peak in the scattering intensity $\mathcal{I}
\left(q\right)$ at a wave number $q=q_\m{max}>0$, which is related to the formation of a MH of
length scale $2\pi/q_\m{max}$, is clearly induced by the addition of antagonistic salt.

The formation of salt-induced MHs has been observed already before in mixtures of water and 
3-methylpyridine by means of SANS \cite{Sadakane2006,Sadakane2007a,Sadakane2007b,
Sadakane2011,Sadakane2014}, and it has been analyzed in terms of a fitting function
\begin{align}
   \mathcal{I}\left(q\right) \simeq 
   \frac{\dps I(0)\left(1+\frac{q^2}{\kappa^2}\right)}
        {\dps \frac{\xi_0^2}{\kappa^2}q^4+\left(\frac{1}{\kappa^2}+\xi_0^2(1-g^2)\right)q^2 +  1}
   \label{eq:Ig2}
\end{align}
with the bulk correlation length $\xi_0$ of the pure, salt-free ($I=0$) solvent, the inverse
Debye length $\kappa=\sqrt{8\pi\ell_BI}$, and a parameter $g^2$ describing solubility contrasts of
the ions \cite{Onuki2004,Bier2012a,Bier2012b}.
It has been shown in Ref.~\cite{Bier2012b} that Eq.~(\ref{eq:Ig2}) is the generic form in the
absence of monopole-dipole interactions, i.e., for the case that the structure formation
is generated by short-ranged interactions and long-ranged monopole-monopole interactions alone.
Attempting to fit Eq.~(\ref{eq:Ig2}) to the SAXS data of the present study of $\m{H_2O}/\m{ACN}$
mixtures leads to unphysical parameters, such as values of $\kappa^2$ which are \emph{negative}
and of wrong magnitude.
Therefore, the intuitively appealing physical picture underlying Eq.~(\ref{eq:Ig2}) of MH
formation due to a competition between short-ranged interactions and Coulomb interactions
amongst the ions does not apply here and one has to find alternatives.
This observation is a clear indication of the importance of monopole-dipole interactions 
between ions and solvent molecules in understanding the formation of salt-induced MHs in 
mixtures of $\m{H_2O}$ and $\m{ACN}$.
Indeed, inspection of Eq.~(\ref{eq:m}) shows that the coefficient $m$ in Eq.~(\ref{eq:Ipade}),
and therefore the position $q_\m{max}=-m/2$ of the maximum of $\mathcal{I}(q)$, is
different from zero only if there are non-vanishing coefficients $c^{(1)}_{ij}$.
The coefficients $c^{(1)}_{ij}$, which originate from Eq.~(\ref{eq:W4}), describe long-ranged 
monopole-dipole interactions.

In contrast to the water/ACN system studied in this work, the small angle scattering patters
from water/3MP mixtures \cite{Sadakane2006,Sadakane2007a,Sadakane2007b,Sadakane2011,Sadakane2014}
lead to physically meaningful parameters using Eq.~(\ref{eq:Ig2}).
This may be caused by the different ratios between specific interactions present in the two systems.
In the general case of non-vanishing monopole-dipole interactions (e.g.\ water/ACN), it is expected
that the scattering patterns $\mathcal{I}(q)$ for binary mixtures of dipolar fluids can
be described by Eq.~(\ref{eq:Ipade}).
In the case of vanishing or negligible monopole-dipole interactions (e.g.\ water/3MP), 
Eq.~(\ref{eq:Ig2}) may apply.
However, so far there is currently no theory available which can a priori predict from common 
solvent properties (Tab.~\ref{Tab:solvents}) whether Eq.~(\ref{eq:Ipade}) or Eq.~(\ref{eq:Ig2}) 
has to be used.


\subsection{\label{Subsec:Spinodal}Spinodal line}

\begin{figure}[!t]
   \includegraphics{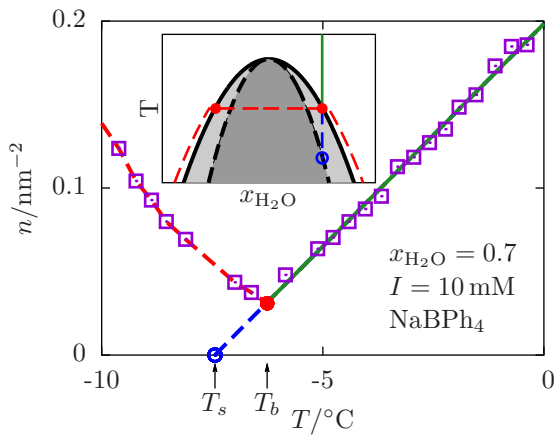}
   \caption{Parameter $n$ of the generic form Eq.~(\ref{eq:Ipade}) of the scattering intensity
            $\mathcal{I}\left(q\right)$ as function of the temperature $T$ for water mole fraction
            $x_\m{H_2O}=0.7$ and ionic strength $I=10\,\m{mM}$ of added $\m{NaBPh_4}$.
            Upon extrapolating (dashed blue line) the linear high-temperature behavior (solid
            green line) one obtains the spinodal temperature $T_s$ (blue circle).
            Below the binodal temperature $T_b$ (red dot) the low-temperature behavior (dashed
            red line) occurs, which corresponds to the phase-separeted system.}
   \label{Fig:2}
\end{figure}

By fitting Eq.~(\ref{eq:Ipade}) to the measured SAXS data one obtains the coefficients $a$, $b$,
$c$, $m$, and $n$ as functions of the solvent composition $x_\m{H_2O}$, the salt type, the ionic
strength $I$, and the temperature $T$.
Inspection of these dependencies led to the observation of $n$ being a linear function of $T$ for
sufficiently high temperatures, as is demonstrated in Fig.~\ref{Fig:2} by the fitted values of $n$
in the range $T > T_b$ (violet squares with a solid green line underneath).
Extrapolation of the linear high-temperature data (dashed blue line) towards $n=0$ (blue circle)
leads to the characteristic temperature $T_s$ and the slope $\mathcal{N}$, by means of which the
dashed blue and the solid green lines in Fig.~\ref{Fig:2} are given as $n(T \geq T_s)=
\mathcal{N}(T-T_s)$.
However, in the low-temperature range $T < T_b$ the fitted values of $n$ (violet squares with a
dashed red line underneath) deviate from the extrapolated linear high-temperature behavior with 
a progressively larger magnitude upon decreasing the temperature.

In order to interpret this finding, one first infers from Eq.~(\ref{eq:Ipade}) that macroscopic 
concentration fluctuations $\mathcal{I}(0)=c/n$ are inversely proportional to $n$ and therefore
maximal at $T=T_b$.
If the measured values of $n$ (violet squares in Fig.~\ref{Fig:2}) followed the linear 
high-temperature trend $n(T \geq T_s)=\mathcal{N}(T-T_s)$ down to $T \searrow T_s$, concentration
fluctuations would diverge ($\mathcal{I}(0)\to\infty$).
This suggests the interpretation of $T=T_s(x_\m{H_2O})$ as the spinodal line in a $T$-$x_\m{H_2O}$
phase diagram (dashed black line in inset of Fig.~\ref{Fig:2}). 
However, except exactly at the critical composition $x_\m{H_2O}=x_{\m{H_2O},c}$, divergence of
concentration fluctuations upon decreasing the temperature is preempted by phase separation, which
takes place at the binodal line $T=T_b(x_\m{H_2O})$ in a $T$-$x_\m{H_2O}$ phase diagram (solid
black line in inset of Fig.~\ref{Fig:2}). 
After phase separation has set in (red dots in inset of Fig.~\ref{Fig:2}), the
distance of the two coexisting phases (dashed red lines in inset of Fig.~\ref{Fig:2})
from the spinodal line increases upon further decreasing the temperature, which leads to a
decrease of the concentration fluctuations $\mathcal{I}(0)$.

\begin{table}[!t]
  \begin{tabular}{|l||c|c|c|c|c|}
      \hline
      $x_\m{H_2O}$       & 0.635 & 0.635 & 0.635 & 0.7   & 0.7   \\
      $I/\m{mM}$         & 0     & 10    & 10    & 0     & 10    \\
      salt               & -     & N     & P     & -     & N     \\
      $T_b/\m{^\circ C}$ & -1.34 & -6.66 & -4.68 & -1.55 & -6.36 \\
      $T_s/\m{^\circ C}$ & -1.36 & -7.25 & -5.15 & -2.26 & -7.44 \\
      \hline
   \end{tabular} 
   \caption{Binodal temperature $T_b$ and spinodal temperature $T_s$ for some systems
            characterized by water mole fraction $x_\m{H_2O}\in\{0.635,0.7\}$, ionic strength 
            $I\in\{0,10\}\,\m{mM}$, and the type of salt (``N'' $\equiv\m{NaBPh_4}$, ``P'' 
            $\equiv\m{PPh_4Cl}$, and ``-'' $\equiv\m{no\ salt}$).}
   \label{Tab:1}
\end{table}

The dependence of the binodal temperature $T_b$ and of the spinodal temperature 
$T_s$ on the composition $x_\m{H_2O}$, on the ionic strength $I$, and on the salt type is
shown in Tab.~\ref{Tab:1}.
For the salt-free ($I=0$) mixture with $x_\m{H_2O}=0.635$ binodal and spinodal temperature 
almost coincide, $T_b \approx T_s$, which is in agreement with the fact that this mole 
fraction is close to the critical concentration $x_{\m{H_2O},c}=0.638$.

\begin{figure}[!t]
   \includegraphics{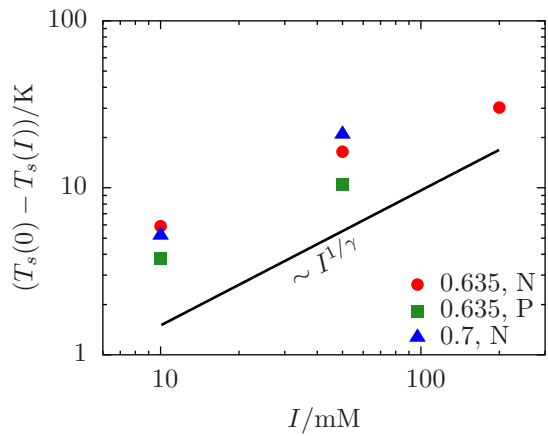}
   \caption{Dependence of the spinodal temperature $T_s$ (blue circle in 
            Fig.~\ref{Fig:2}), on the ionic strength $I$ for systems with water mole fraction
            $x_\m{H_2O}\in\{0.635,0.7\}$ and some antagonistic salts (``N'' $\equiv\m{NaBPh_4}$
            and ``P'' $\equiv\m{PPh_4Cl}$).
            The spinodal temperature $T_s(0)$ for pure, salt-free ($I=0$) mixtures is
            displayed in Tab.~\ref{Tab:1}.
            The scaling relation $T_s(0)-T_s(I)\sim I^{1/\gamma}$ with
            the universal critical exponent $\gamma$ can be justified by means of general
            arguments (see main text).}
   \label{Fig:3}
\end{figure}

The dependence of the spinodal temperature $T_s(I)$ on the ionic strength $I$ is displayed in
Fig.~\ref{Fig:3} for water mole fraction $x_\m{H_2O}\in\{0.635,0.7\}$ and antagonistic salts
(``N'' $\equiv\m{NaBPh_4}$ and ``P'' $\equiv\m{PPh_4Cl}$). 
Realizing that the denominator on the right-hand side of Eq.~(\ref{eq:n}) measures the macroscopic
concentration fluctuations of the salt-free mixture, one expects the scaling behavior
\begin{align}
   \big(1-c^{(0)}_{AA}\big)\big(1-c^{(0)}_{BB}\big) - \big(c^{(0)}_{AB}\big)^2
   \sim (T_s(0)-T)^\gamma
   \label{eq:saltfreefluc}
\end{align}
in the temperature range $T<T_s(0)$, where $\gamma \approx 1.2372$ is the well-known critical
exponent of order parameter fluctuations of the 3D-Ising universality class \cite{Pelissetto2002}.
By definition, $n$ vanishes at $T=T_s(I)$, and hence, from Eqs.~(\ref{eq:n}) and 
(\ref{eq:saltfreefluc}), one infers $(T_s(0)-T_s(I))^\gamma \sim I$, i.e.,
\begin{align}
   T_s(0)-T_s(I) \sim I^{1/\gamma}.
   \label{eq:Ts0TsI}
\end{align}
This scaling behavior is reasonably well confirmed by the experimental data in Fig.~\ref{Fig:3}.


\subsection{\label{Subsec:Critical}Critical-like behavior}

Upon approaching the critical point, well-known critical behavior occurs, e.g., the
divergence of the concentration fluctuations $\mathcal{I}(0)$ and of the bulk correlation length
$\xi$ according to power laws with universal critical exponents \cite{Pelissetto2002}.
Moreover, the same critical-like behavior can be expected to occur upon approaching the spinodal
line $T=T_s(x_\m{H_2O},I)$ anywhere, i.e., not only at the critical point.

\begin{figure}[!t]
   \includegraphics{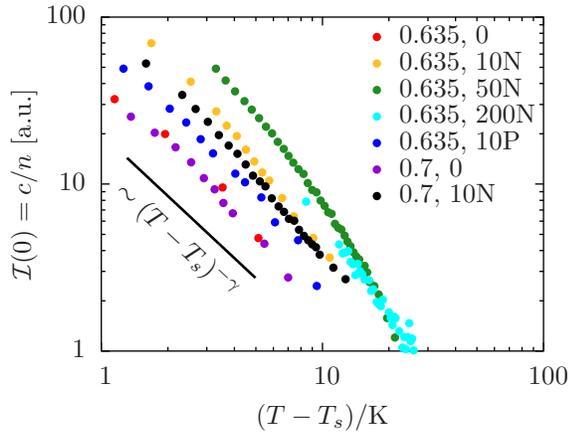}
   \caption{Macroscopic concentration fluctuations $\mathcal{I}(0)$ for water mole fraction
            $x_\m{H_2O}\in\{0.635,0.7\}$, ionic strength $I\in\{0,10,50,200\}\,\m{mM}$, and
            some salts (``N'' $\equiv\m{NaBPh_4}$ and ``P'' $\equiv\m{PPh_4Cl}$) as function
            of the temperature difference $T-T_s$ from the spinodal.
            Close to the spinodal universal critical-like behavior $\mathcal{I}(0)\sim
            (T-T_s)^{-\gamma}$ with the universal critical exponent $\gamma$ is observed.}
   \label{Fig:4}
\end{figure}

Figure~\ref{Fig:4} displays $\mathcal{I}(0)=c/m$ as function of the temperature difference $T-T_s$
from the spinodal line for solvent composition $x_\m{H_2O}\in\{0.635,0.7\}$, ionic strength
$I\in\{0,10,50,200\}\,\m{mM}$ and antagonistic salts (``N'' $\equiv\m{NaBPh_4}$ and ``P'' 
$\equiv\m{PPh_4Cl}$). 
At small temperature distances $T-T_s$ inside the one-phase region of the phase diagram,
i.e., for $T>T_b$, the expected universal scaling behavior $\mathcal{I}(0)\sim(T-T_s)^{-\gamma}$
with the universal critical exponent $\gamma\approx 1.2372$ (Ref.~\cite{Pelissetto2002}) is 
confirmed for all systems.

\begin{figure}[!t]
   \includegraphics{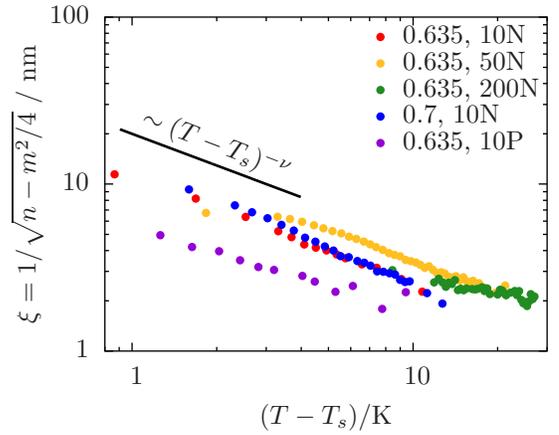}
   \caption{Correlation length $\xi$ for water mole fraction $x_\m{H_2O}\in\{0.635,0.7\}$, ionic
            strength $I\in\{10,50,200\}\,\m{mM}$, and some salts (``N'' $\equiv\m{NaBPh_4}$ 
            and ``P'' $\equiv\m{PPh_4Cl}$) as function of the temperature difference $T-T_s$ from
            the spinodal.
            Close to the spinodal universal critical-like behavior $\xi\sim(T-T_s)^{-\nu}$ with
            the universal critical exponent $\nu$ is observed.}
   \label{Fig:5}
\end{figure}

Similarly, Fig.~\ref{Fig:5} displays $\xi=1/\sqrt{n-m^2/4}$ as function of the temperature
difference $T-T_s$ from the spinodal line for solvent composition $x_\m{H_2O}\in\{0.635,0.7\}$,
ionic strength $I\in\{10,50,200\}\,\m{mM}$ and antagonistic salts (``N'' $\equiv\m{NaBPh_4}$ and
``P'' $\equiv\m{PPh_4Cl}$).
Again, the expected universal scaling behavior $\xi\sim(T-T_s)^{-\nu}$ with the universal
critical exponent $\nu\approx 0.6301$ (Ref.~\cite{Pelissetto2002}) is found.

These results show the consistency of the interpretation of $T_s$ as the spinodal temperature,
with respect to which critical-like universality is expected to occur.
Moreover, the critical-like behavior found for the present systems all belongs to the 3D-Ising
universality class.
Hence, adding the antagonistic salts $\m{NaBPh_4}$ or $\m{PPh_4Cl}$ to $\m{H_2O}+\m{ACN}$ mixtures
does not alter the universality class. 


\subsection{\label{Subsec:StrucMH}Structure of microheterogeneities}

\begin{figure}[!t]
   \includegraphics{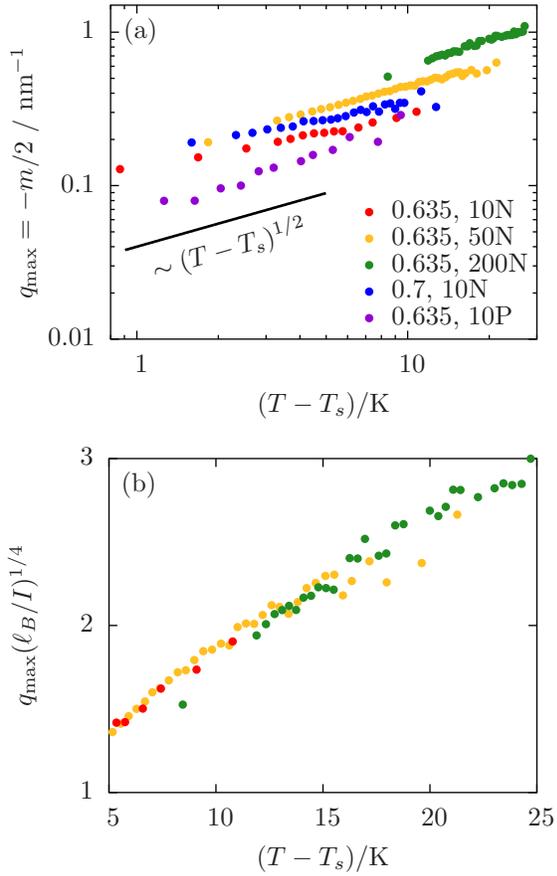}
   \caption{Dependence of the wave number $q_\m{max}$ of the maximum of the scattering
            intensity $\mathcal{I}\left(q\right)$ (Fig.~\ref{Fig:1}) as function of the 
            temperature difference $T-T_s$ from the spinodal for water mole fraction
            $x_\m{H_2O}\in\{0.635,0.7\}$, ionic strength $I\in\{0,10,50,200\}\,\m{mM}$, and 
            some salts (``N'' $\equiv\m{NaBPh_4}$ and ``P'' $\equiv\m{PPh_4Cl}$).
            The wave number $q_\m{max}$ is related to the characteristic length scale
            $2\pi/q_\m{max}$ of the MH.
            Panel (a) confirms the scaling $q_\m{max} \sim (T-T_s)^{1/2}$ for small
            temperature differences $T-T_s$ from the spinodal derived in the main text.
            The collapse of the data points onto one curve in panel (b) shows the scaling
            $q_\m{max} \sim I^{1/4}$ with the ionic strength $I$ for sufficiently large
            temperature differences $T-T_s$ from the spinodal.}
   \label{Fig:6}
\end{figure}

As already mentioned after Eq.~(\ref{eq:n}), the scattering intensity $\mathcal{I}\left(q\right)$
exhibits a maximum at wave number $q=q_\m{max}=-m/2$ with a peak width of half height $2/\xi$.
This maximum is related to a characteristic wave length $2\pi/q_\m{max}$ of concentration
fluctuations, which decay on the scale of the correlation length $\xi=1/\sqrt{n-m^2/4}$.
Since $\xi\sim(T-T_s)^{-\nu},\nu\approx0.6301$, (Fig.~\ref{Fig:5}) and $n\sim T-T_s$ 
(Fig.~\ref{Fig:2}) for $T\searrow T_s$, one expects 
\begin{align}
   q_\m{max} = -\frac{m}{2} 
             = \sqrt{n-\frac{1}{\xi^2}} 
             \simeq \sqrt{n} 
             = \sqrt{\mathcal{N}} (T-T_s)^{1/2}.
   \label{eq:qmaxT}
\end{align}
This scaling of $q_\m{max}$ with respect to $T-T_s$ is confirmed in Fig.~\ref{Fig:6}(a) for
solvent composition $x_\m{H_2O}\in\{0.635,0.7\}$, ionic strength $I\in\{10,50,200\}\,\m{mM}$ and 
antagonistic salts (``N'' $\equiv\m{NaBPh_4}$ and ``P'' $\equiv\m{PPh_4Cl}$).

It is found empirically, that, given composition $x_\m{H_2O}$ and salt type, the quantity
$q_\m{max}/I^{1/4}$ depends not on the ionic strength $I$ for sufficiently large temperature
differences $T-T_s$ from the spinodal, which is shown in Fig.~\ref{Fig:6}(b) for the case
$x_\m{H_2O}=0.635$ and $\m{NaBPh_4}$.
Consequently, at sufficiently high temperatures $T$ above the spinodal temperature $T_s$, the
wave number at the peak position $q_\m{max}$ scales as $q_\m{max} \sim I^{1/4}$.


\section{\label{Sec:conclusion_summary}Conclusion and Summary}

All $\m{H_2O}/\m{ACN}$ mixtures with different concentrations of the two antagonistic salts 
$\m{NaBPh_4}$ and $\m{PPh_4Cl}$ exhibit MHs with characteristic length scales in the 
$\m{nm}$-regime.
It turned out that MH formation in these systems cannot be attributed to monopole-monopole 
interactions between the ions alone, but that monopole-dipole interactions between ions and solvent
molecules are necessary for a quantitative understanding.
By taking into account electric monopole-dipole interactions, a generic form of the SAXS pattern 
$\mathcal{I}\left(q\right)$ has been derived (Sec.~\ref{Subsec:GenFormScatIn}).
Using Eq.~(\ref{eq:Ipade}), the experimental SAXS data can be quantitatively reproduced by fitting
(Fig.~\ref{Fig:1}).
The resultant quantities are: 
The amplitude of the macroscopic concentration fluctuations (Fig.~\ref{Fig:4}), the bulk
correlation length (Fig.~\ref{Fig:5}), and the characteristic periodicity of the MH 
(Fig.~\ref{Fig:6}a).
In contrast to the parameters extracted by fitting Eq.~(\ref{eq:Ig2}), i.e., the standard model
for MHs, those obtained by fitting Eq.~(\ref{eq:Ipade}) are all physically meaningful.
Detailed analysis showed, that their temperature-dependence is governed by the distance from the
spinodal line $T=T_s(x_\m{H_2O},I)$ in the phase diagram (Fig.~\ref{Fig:2}).
Upon adding salt, the spinodal line shifts to lower temperatures (Fig.~\ref{Fig:3}).

A physical understanding of the mechanisms leading to MH formation caused by monopole-dipole 
interactions is obtained by analysis of Eq.~(\ref{eq:m}).
Its relevance is given by the relation of $m$ to the wave number $q_\m{max}=-m/2$ of the MH.
The coefficients $c^{(1)}_{ij}$ in Eq.~(\ref{eq:m}) originate exclusively from the long-ranged part
of the monopole-dipole interaction (Eq.~(\ref{eq:W4})).
Typically, induced dipoles are much weaker than permanent ones.
Therefore, expression $c^{(1)}_{\oplus\oplus}+2c^{(1)}_{\oplus\ominus}+c^{(1)}_{\ominus\ominus}$ in
Eq.~(\ref{eq:m}) can be neglected.
The dominant last term in the bracket of Eq.~(\ref{eq:m}) may be rewritten in order to obtain
\begin{align}
   q_\m{max} \approx M
   \left(\begin{array}{c}
      T^{(0)}_A \\ 
      T^{(0)}_B
   \end{array}\right)
   \cdot
   \left(\begin{array}{cc}
      1-c^{(0)}_{AA} &  -c^{(0)}_{AB} \\
      -c^{(0)}_{AB}  & 1-c^{(0)}_{BB}      
   \end{array}\right)^{-1}
   \left(\begin{array}{c}
      T^{(1)}_A \\ 
      T^{(1)}_B
   \end{array}\right).
   \label{eq:mmod}   
\end{align}
The inverse matrix in Eq.~(\ref{eq:mmod}) corresponds to the partial structure factors 
$\underline{\underline{S}}^{(0)}$ of the pure, salt-free mixture at wave number $q=0$.
Hence, due to Yvon's equation \cite{Hansen1986}, it is proportional to the integral of the 
density-density correlation matrix $\underline{\underline{\widehat{G}}}^{(0)} = \varrho 
\underline{\underline{S}}^{(0)}$.
Therefore, Eq.~(\ref{eq:mmod}) expresses the scenario where MHs with $q_\m{max}\not=0$ 
originate from a coupling (represented by $\underline{\underline{\widehat{G}}}^{(0)}$) of 
short-ranged interactions (represented by $\underline{T}^{(0)}$)
and long-ranged monopole-dipole salt-solvent interactions (represented by 
$\underline{T}^{(1)}$).
It is important to note that 
$T^{(k)}_i := c^{(k)}_{i\oplus}+c^{(k)}_{i\ominus}, i\in\{A,B\},k\in\{0,1\}$, 
is the \emph{sum} of cation-solvent and anion-solvent contributions.
In contrast, the scenario described in Refs.~\cite{Onuki2004,Bier2012a,Bier2012b} is based on the 
\emph{differences} between the cation-solvent and anion-solvent interactions. 

Based on these formal results, the following picture emerges: 
A competing mechanism between charge fluctuations and their monopole-dipole interaction leads to
the formation of MHs.
For an antagonistic salt, the ion species are preferably solvated by different
solvent components.
The difference is generated by short-ranged interactions, leading to short-ranged
correlations only.
The preference of antagonistic ions for different solvent components leads to solvation-induced
short-ranged charge density fluctuations, which give rise to long-ranged monopole-dipole
interactions.
These long-ranged interactions are strongest for the more polar solvent component.
The relative strength of short-ranged interactions and long-ranged monopole-dipole 
interactions determines the characteristic length scale $2\pi/q_\m{max}$ of these MHs: 
The stronger the long-ranged monopole-dipole interaction, the larger $q_\m{max}$ (Eqs.~(\ref{eq:m})
or (\ref{eq:mmod})), i.e., the smaller the characteristic length scale of the MH.

Recently, it has been argued that the absence of MH in $\m{H_2O}$/$\m{3MP}$ mixtures
with simple inorganic salts \cite{Sadakane2006,Sadakane2007a,Sadakane2007b,Sadakane2011,
Sadakane2014} may be caused by similar anion and cation sizes \cite{Bier2012b}.
However, there the considered inorganic salts are not antagonistic.
Within the picture proposed above, the absence of MH can therefore also be understood by a
different mechanism:
The absence of charge fluctuations leads to vanishing long-ranged monopole-dipole
interactions.
Accordingly, in the present study, $\m{H_2O}$/$\m{ACN}$ mixtures with $I\in\{10,50\}\,\m{mM}$
$\m{NaCl}$ exhibit no MH.
This observation is in agreement with the findings of Takamuku et al.\ \cite{Takamuku2001a,
Takamuku2001b, Takamuku2007a, Takamuku2007b, Takamuku2007c, Haramaki2013}.

In summary, salt-induced MHs in $\m{H_2O}$/$\m{ACN}$ mixtures with the antagonistic salts
$\m{NaBPh_4}$ or $\m{PPh_4Cl}$ have been systematically studied by SAXS.
A detailed analysis of these data suggests that these MHs are generated by a 
competition of short-ranged interactions and long-ranged electrostatic monopole-dipole
interactions.
Besides being consistent with the present and previous experimental results, this picture offers a
first explanation for the occurrence of characteristic length scales of MHs.

In chemical reactions, microheterogeneous solvent structures can influence their catalytic 
activity.
These processes in a macroscopically homogeneous liquid phase can be described as phase transfer
or interfacial reactions at domain boundaries \cite{Holloczki2017}.
Therefore, the possibility of MH formation with controlled length scales using near critical 
solvent mixtures with ionic impurities might offer an attractive approach to tune catalytic
reactions.
Being partly of electrostatic origin, salt-induced {MHs} in fluids may be used for pattern
formation at interfaces.
Here, the electrodes could serve to control the pattern's size and morphology.
Detailed studies on such salt-induced MH at interfaces are planned for future investigations.


\begin{acknowledgments}
The authors acknowledge Henning Weiss, Stefan Geiter, Xilin Wu, and Gunner Kircher from MPI-P for
their help with SAXS measurements and sample preparation as well as Akira Onuki for useful
comments.
J.~M.\ and M.~M.\ acknowledge the MAINZ Graduate School of Excellence, funded through the 
Excellence Initiative (DFG/GSC 266) for financial support.
H.~L.\ was supported by the China Scholarship Council.
\end{acknowledgments}


\end{document}